\documentclass[twocolumn,secnumarabic,amssymb, nobibnotes, aps, prd]{revtex4-1}
\usepackage{graphicx}%
\usepackage{siunitx}

\begin{document}
	
\title{First Direct-Drive Measurements of Laser--Imprint--Induced Shock Velocity Nonuniformities}%
	
\author{J. L. Peebles$^{1}$, S. X. Hu$^{1}$, W. Theobald$^{1}$, V. N. Goncharov$^{1}$, N. Whiting$^{1}$, P. M. Celliers$^{2}$, \\ S. J. Ali$^{2}$, G. Duchateau$^{3}$, E. M. Campbell$^{1}$, T. R. Boehly$^{1}$ and S. P. Regan$^{1}$}%
\affiliation{$\mathit{^1}$Laboratory for Laser Energetics, University of Rochester, 250 E. River Road, Rochester, NY 14623, USA}
\affiliation{$\mathit{^2}$Lawrence Livermore National Laboratory, 7000 East Avenue, Livermore, CA 94550, USA}
\affiliation{$\mathit{^3}$University Bordeaux-CNRS-CEA, Centre Lasers Intenses et Applications, UMR 5107, 33405 Talence, France}

\date{\today}

\begin{abstract}
Perturbations in the velocity profile of a laser-ablation-driven shock wave seeded by speckle in the spatial beam intensity (i.e., laser imprint) have been measured for the first time.  Direct measurements of these velocity perturbations were recorded using a two-dimensional (2-D) high-resolution velocimeter probing plastic material shocked by 100-ps picket laser pulse from the OMEGA Laser System.  The measured results for experiments with one, two, and five overlapping beams incident on target clearly demonstrate a reduction in long-wavelength ($ > 25\ \mu$m) perturbations  with an increasing number of overlapping laser beams, consistent with theoretical expectations. These experimental measurements are crucial to validate radiation-hydrodynamics simulations of laser imprint for laser direct drive inertial confinement fusion research, since they highlight the significant (factor of three) underestimation of the level of seeded perturbation when the microphysics processes for initial plasma formation such as multiphoton ionization are neglected.
\end{abstract}
	
\pacs{numbers!}
	
\maketitle
\section{Introduction}
Laser direct drive (LDD) inertial confinement fusion (ICF) involves the direct laser irradiation of a plastic spherical shell target containing a thin layer of cryogenic, thermonuclear fuel (i.e., deuterium (D) and tritium(T)) with symmetrically-arranged, high-intensity, overlapping laser beams \cite{Craxton}. The resulting laser-ablation process launches a spherical shock wave into the target and accelerates the shell inward via the rocket effect. Nonuniformities in the laser drive due to laser speckle and beam-to-beam intensity variations can seed hydrodynamic instabilities such as the Richtmyer--Meshkov (R-M) and Rayleigh--Taylor (R-T) instabilities \cite{Bodner}. The physical energy transfer of the laser-intensity modulations to the shock front, called laser imprint, depends strongly on the initial plasma formation \cite{Bourgeade}. The resulting coronal plasma provides a physical standoff between the laser deposition region and the ablation surface in a few hundred picoseconds and the efficiency of the laser imprint process drops to zero. Hydrodynamic instabilities seeded by laser imprint as well as mass perturbations in the target could adversely affect the compression of the imploding shell and the DT nuclear fusion yield (i.e., D + T $\rightarrow$ n + $\alpha$) achieved at stagnation ~\cite{Remington, Glendinning, Azechi, Azechi2, Cherfils, Smalyuk, Smalyuk2,Hu2,Michel,Hu4}. For the optics community, it is also a concern that laser speckle interacting with material defects can cause damage to optical materials \cite{Carr}. Thus, understanding laser imprinting process is of great importance for LDD ICF research and laser damage to optical materials.

Several mitigation methods for laser-imprint have been developed. The principal efforts involve smoothing the spatial intensity profile of each laser beam using distributed phase plates (DPPs)~\cite{Kato,Lin}, smoothing by spectral dispersion (SSD)~\cite{Skupsky,Regan,Regan2,Lehmberg} and polarization smoothing using distributed polarization rotators (DPRs)\cite{Boehly2} in glass lasers and by using induced spatial incoherence (ISI)~\cite{Lehmberg,Lehmberg2,Deniz} in excimer lasers such as KrF. The mitigation of laser imprint using SSD bandwidth has been shown to lead to a twofold increase in implosion performance\cite{Hu2,Michel,Hu4}. While these techniques have significantly improved laser uniformity and therefore lowered imprint-seeded R-M and R-T instabilities, modeling laser imprint during the initial plasma formation is extremely challenging and is at the forefront of LDD research. These simulations require quantitative measurements of laser imprinting in order to be calibrated. Previous experiments characterizing laser imprint relied on R-M and R-T instabilities to amplify imprint seeds to detectable levels and then characterized the larger features using x-ray radiography ~\cite{Kalantar,Kalantar2,Kalantar3,Depierreux,Obenschain,Karasik,Mostovych,Fujioka,Michel}. However, to most accurately characterize the development and significance of laser imprint, one needs a direct measurement of the shock-velocity perturbation produced by laser perturbations with no intermediate hydrodynamic processes.

While intensity modulations in the incident laser and development of R-M and R-T instabilities in a shocked shell are well understood and simulated processes, the coupling between the two during the plasma build up period  is not. Efforts to understand and mitigate this complex laser imprint phase have been performed using 2-D radiation-hydrodynamics simulations ~\cite{Hu,Hu2,Hu3,Olazabal-Loumé,Masse}. These simulations routinely approximate the initial plasma formation and laser imprint by using an experimentally derived intensity profile representative of the driving laser to seed the hydrodynamic instabilities in a pre-ioinized plasma. This approximation does not include the initial material breakdown via multiphoton ionization. Non-linear absorption effects during the initial plasma formation must be accounted for to accurately capture the laser-target interaction \cite{Bourgeade}. Experiments presented in this manuscript demonstrate that the current approach, which does not include the initial plasma formation, grossly under predicts the magnitude of perturbation imprinted by the laser on target by a factor of three. This level of under prediction of the laser imprint could significantly reduce the neutron yield and compressed areal density of the implosion \cite{Hu2,Michel}.   
	
\section{Experimental Setup and the OHRV Diagnostic}

\begin{figure}[b]
	\includegraphics[width=\columnwidth]{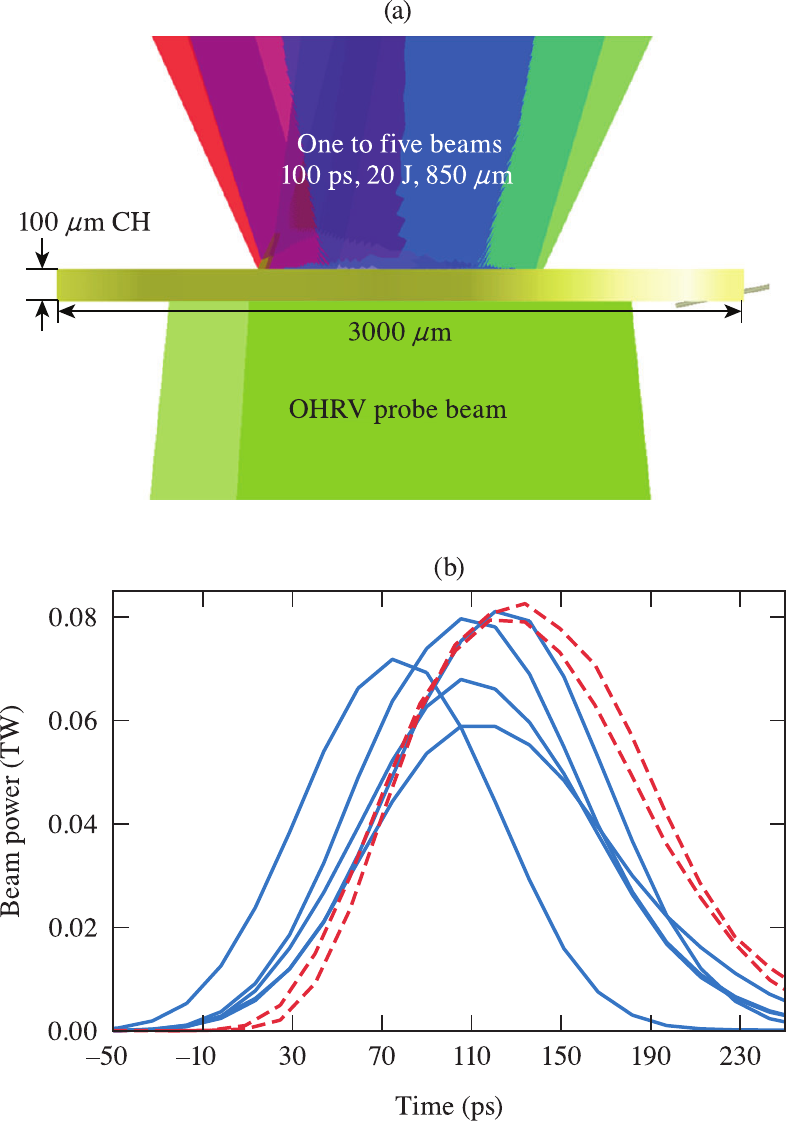}
	\caption{(a) Experimental setup: one, two, or five beams delivering 20 J in a 100-ps picket are incident on a 100-$\mu$m CH polymer disk. A shock is generated and propagates through the medium, where it is probed by the OHRV beam 0.9 ns later. (b) Relative timing for beams in the five-beam experiment (solid blue) and the two-beam experiment (dashed red). In the five-beam experiment, one beam was roughly 20 ps early compared to the others.}
\end{figure}
	
Presented are the first direct experimental observations of shock-velocity fluctuations induced by laser imprint and the reduction of imprint produced by the interference of multiple overlapping beams. This is the most direct measurement of the magnitude of laser imprint and provides critical experimental insights to validate the simulation and theory of laser imprinting. The experiment was carried out on the 60-beam, 30-kJ, 351-nm OMEGA Laser System \cite{Boehly}. As shown in Fig. 1(a), the experiment consisted of a 3-mm outer diameter, 100-$\mu$m-thick planar disk of CH polymer irradiated by one, two, or five of the OMEGA beams using a 100-ps ``picket'' pulse containing approximately 20 J of total energy. These beams were spatially smoothed using DPRs with SG4, 850-$\mu$m DPPs where 95\% of the beam's energy is contained within a diameter of 850 $\mu$m with a 500-$\mu$m flat top region. The limiting cases of no SSD bandwidth and overlapping beams are examined herein. The intensity of the overlapped beams was approximately $10^{13}\ \mathrm{W/cm^2}$ for all cases. This configuration was chosen to replicate the effect of picket pulses used in direct-drive pulse shapes since pickets cause most of the initial laser-imprint seed. By using a single picket, the shock is unsupported and perturbation growth from R--M and R--T instabilities are minimized. Each beam was incident at a \ang{21.4} angle relative to target normal.

The target was placed so that the rear surface faced directly toward the OMEGA high-resolution velocimeter (OHRV) diagnostic \cite{Celliers}. The OHRV detects perturbations in shock velocity using a push--pull VISAR (velocity interferometer system for any reflector) system \cite{Hemsing} by using the interference pattern produced by two short, 2-ps, 395-nm laser pulses separated by a specified delay. The fringe pattern is a 2-D phase map snapshot, which is proportional to the velocity of the shock surface. The phase map is deconstructed and post-processed using methods based on work by Celliers \textit{et al.} ~\cite{Celliers, Hemsing, Erskine, Erskine2, Ghiglia, Ghiglia2}. The spatial resolution, limited by the internal optical system,  is 2 to 3 $\mu$m in the target plane. The interferometers were set up to resolve velocity changes of 3.8 km/s per fringe shift. For this experiment the OHRV probe was timed 0.9 ns after the start of the drive beams to provide adequate time for the CH target material to recover from ‘optical blanking’ due to ionization in the target \cite{Theobald}.


\begin{figure}[b]
	\includegraphics[width=\columnwidth]{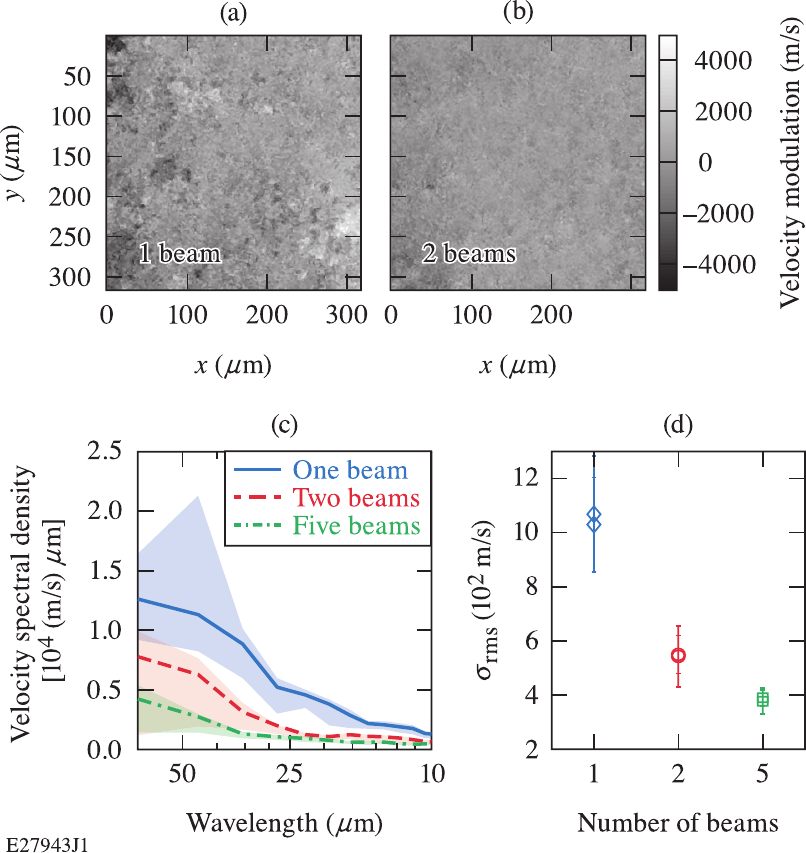}
	\caption{2-D velocity modulation maps taken by OHRV over a 315 x 315 $\mu$m region in the uniformly driven portion of the target plane for (a) one and (b) two beam configurations. Qualitatively one can see significantly more perturbations in the one beam case across all modes. (c) Radially averaged velocity spectral information inferred from OHRV measurements for (a), (b), and five beams. (95\%) confidence intervals are represented by the shaded region for each spectrum. The single beam case indeed shows more energy in perturbations across all modes when compared to the two or five beam cases. (d) $\sigma_\mathrm{rms}$ for modes of interest in all shots for each experimental case demonstrate repeatability of the data}
\end{figure}

Two examples of the post-processed OHRV data from the experiment are shown in Fig. 2. These are 2-D velocity modulation maps over a 315 $\times$ 315-$\mu$m region in the uniformly driven portion of the target plane for the one-beam [Fig. 2(a)] and two-beam [Fig. 2(b)] cases. One can qualitatively see the single-beam case contains more significant perturbations than the two-beam case. To quantify the velocity nonuniformities a 2-D Fourier transform is taken of the image, normalized to the image size, and azimuthally averaged for each mode. The Euclidean norm taken of this spectrum to construct a 1-D velocity spectral density of the image with respect to mode size. Fig. 2(c) compares the velocity spectra for one shot from each of the experimental cases, showing that the one-beam case has significantly more energy in velocity perturbations across all measurable wavelengths. Confidence intervals (95\%) for the velocity spectra were calculated by measuring the variance of the spectra of individual lineouts along the vertical and horizontal axes across the entire image. The measured $\sigma_\mathrm{rms}$ for the processed data was approximately 1030 $\pm$ 174, 550 $\pm$ 70 and 390 $\pm$ 32 m/s for the one-, two-, and five-beam cases, respectively. The $\sigma_\mathrm{rms}$ is shown in Fig. 2(d) for each experimental case, which was duplicated to demonstrate the repeatability of the diagnostic and experiment. Error was calculated using the variance of the $\sigma_\mathrm{rms}$ across the images along both axes.

It was shown by Smalyuk \textit{et al.} \cite{Smalyuk3} that the inclusion of additional overlapping beams reduces the modulation of the shock front in laser-accelerated foils that experienced RT growth. The anticipated reduction in the $\sigma_\mathrm{rms}$ is a factor of $\sqrt{N}$, where $N$ is the number of beams. If one applies this factor to the one beam case ($\sigma_\mathrm{rms} \approx 1030$) then a $\sigma_\mathrm{rms} \approx 730$ and $460$ is expected for the two- and five-beam cases respectively. In the present experiment however, the reduction of velocity modulation was greater than anticipated when moving from one to two beams and less than expected moving from two to five beams. A possible cause for reduced imprint mitigation in the five-beam case is slight beam mistiming. For both shots with five beams the laser pulse diagnostic recorded that one beam arrived approximately 20 ps earlier than the others [Fig. 1(b)]; the resulting imprinted velocity perturbations were therefore stronger because of this single beam and the apparent mitigation lessened.

\section{2D \textit{DRACO} Simulations Underestimate Imprinted Modulations}
\begin{figure}[b]
	\includegraphics[width=\columnwidth]{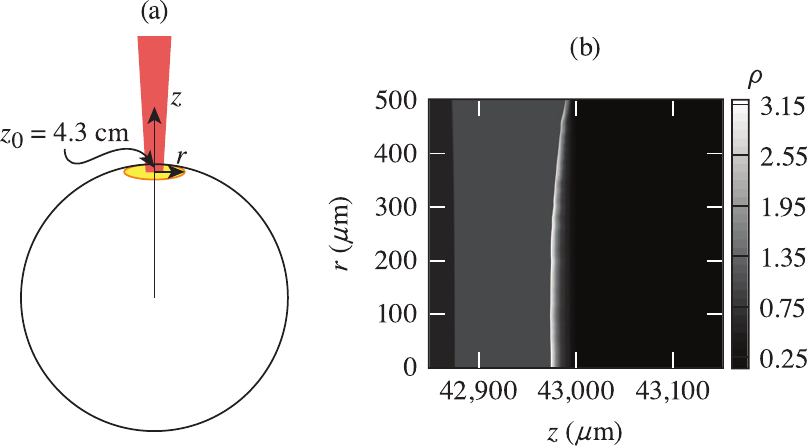}
	\caption{(a) A cartoon of the 2D cylindrical geometry created to approximate a planar solution for the DRACO simulations. (b) A zoomed in image at the large sphere radius where the beam interacts with the target. One can see that the target surface (light blue) approximates a planar surface very closely as the beam along the z-axis pushes into it.}
\end{figure}
These experiments provide valuable quantitative data that are used to calibrate hydrodynamic simulations by providing direct measurements of shock-velocity perturbations produced by intensity nonuniformities without intermediate physical processes of the hydrodynamic instabilities. Initial comparison between experimental results and simulations began by simulating the simplest single beam experiment. Simulations of this experiment were performed using the Lagrangian version of the 2-D radiation-hydrodynamic code \textit{DRACO} \cite{Radha}. The simulations were constructed in an \textit{r--z} geometry shown in Figure 3 (i.e., \textit{r} lies in the target plane and \textit{z} lies in the distance traveled direction orthogonal to \textit{r}). In order to conduct appropriate simulations for planar experiments using an \textit{r--z} geometry, a small section of a large radius surface was simulated \cite{Hu,Hu5}. Open boundary conditions were used for the \textit{r}-axis and reflective boundary conditions for the \textit{z}-axis. The grids included 400x1000 zones for \textit{z} and \textit{r} coordinates, respectively, which generated converged results. These simulations used 3-D ray tracing, a flux-limited thermal conduction model and a pre-ionized plasma approximation to start the simulation.

The laser-intensity profile was constructed using a sum of sinusoidal modes whose amplitudes are determined from the measured intensity profile for a beam using an 850-$\mu$m DPP on the OMEGA Laser System \cite{Epstein,Hu4}. A random numerical seed is applied to these modes in order to randomize their spatial distribution in the simulation. An example of such an intensity profile is shown in Fig. 4(a), which includes modes up to $\ell = 200$ ($\lambda \approx 10\ \mu$m). Modes higher than $\ell = 200$ were found to cause numerical noise in the simulation. Although the OHRV can resolve modes beyond this, down to 2-3 $\mu$m, comparisons were focused on modulations down to $\lambda \approx 10\ \mu$m to have a fair comparison between simulations and experimental data. Fig. 4(b) illustrates the simulated trajectory of the shock front as a function of time and space (\textit{r,z}). The spatially varying shock position is extracted from this trajectory at 900 ps (consistent with the probe time of the experiment).

\begin{figure}[b]
	\includegraphics[width=\columnwidth]{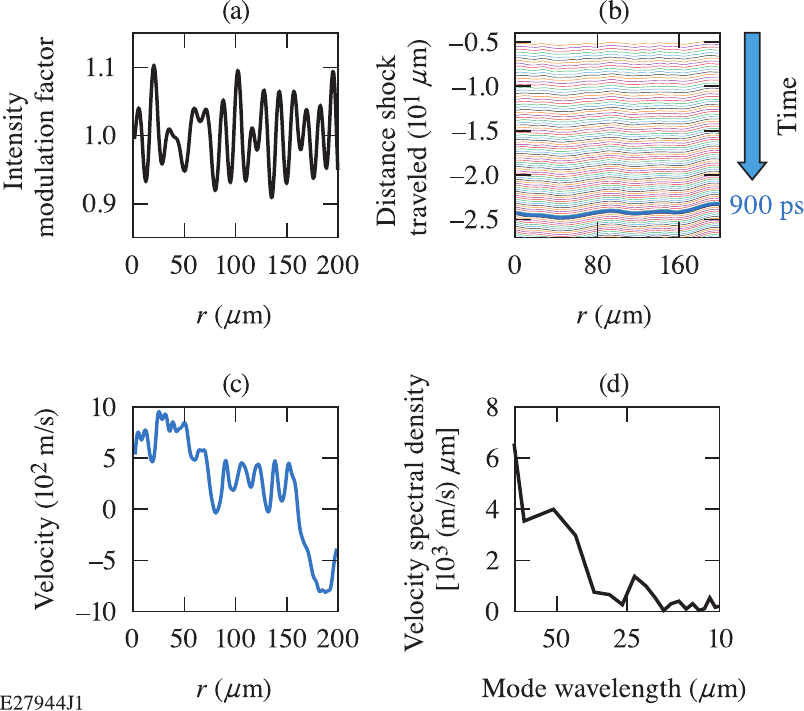}
	\caption{(a) Example laser intensity profile input to \textit{DRACO} for the one beam case. (b) Simulated position of the shock front as time progresses. (c) The calculated velocity of the shock front. (d) The normalized velocity spectrum is calculated from this velocity profile.}
\end{figure}
\begin{figure}[]
	\includegraphics[width=\columnwidth]{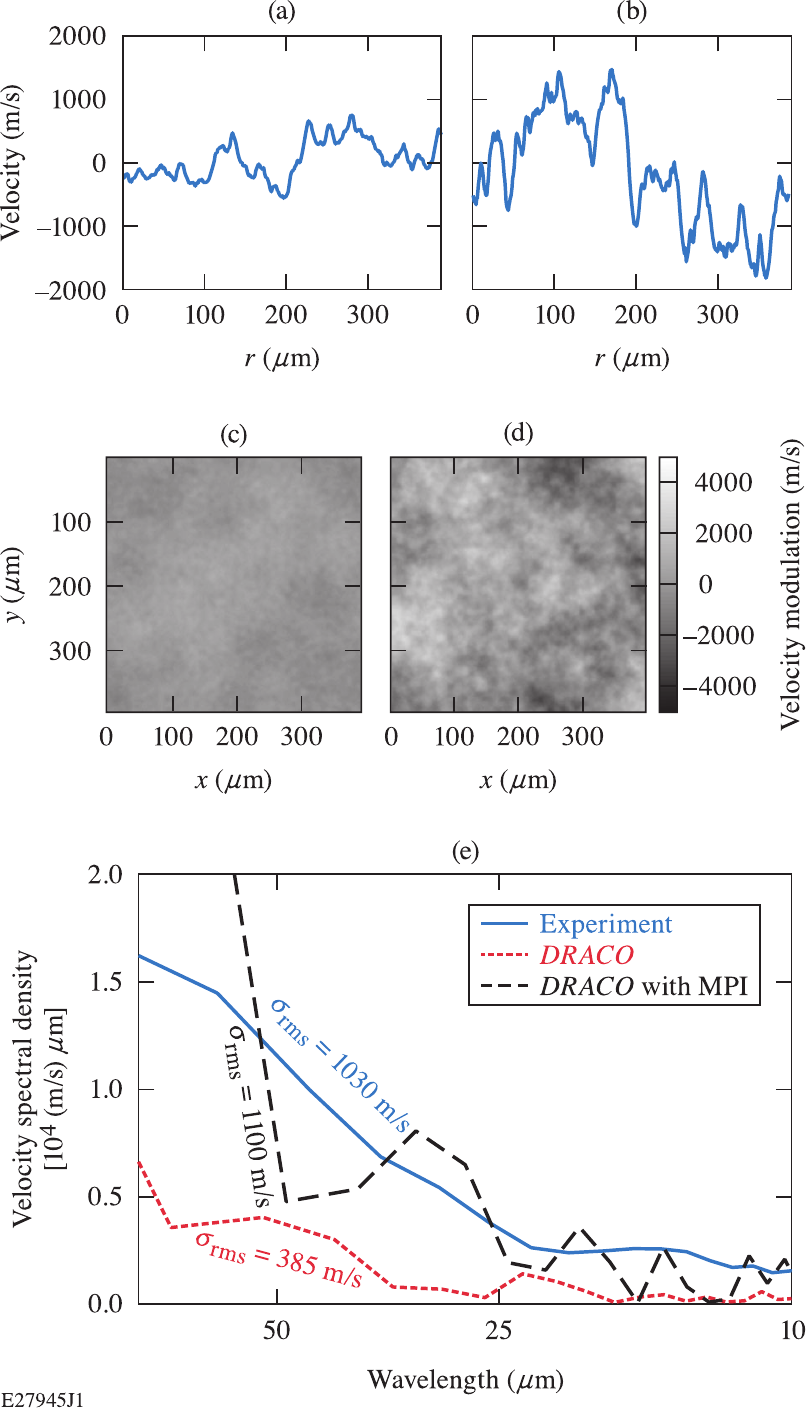}
	\caption{[(a),(c)] Lineout and extrapolated 2-D velocity modulation map of the \textit{DRACO}-calculated shock front for the one beam case and [(b),(d)] one beam with MPI accounted for. One can see that the simulation without MPI drastically underestimates the degree of imprint, especially when compared to Fig. 2(a). (e) A comparison of the velocity spectra from \textit{DRACO} (red and black) to the experimental data (blue) for the one beam case.}
\end{figure}

Since \textit{DRACO} simulations are performed in 2-D \textit{r--z} geometry, the resulting velocity profile is in 1-D (along the \textit{r} axis) as shown in Fig. 4(c). The simulated velocity profile must therefore be converted to be comparable to the 2-D experimental data. The normalization and velocity spectral density processes applied to experimental data are also applied to the simulated velocity profile [Fig. 4(d)]. Using the resulting velocity spectrum, a 2-D synthetic velocity image is created by reversing the process, albeit starting from a position of less information in two aspects. To start, the 1-D velocity spectrum is distributed in a 2-D map; however, if one were to simply take the inverse Fourier transform of this map an image would be produced with perfect spatial symmetry (an unphysical appearing result). Therefore Gaussian white noise is applied spatially to the 2-D map such that when the azimuthal average is taken the original spectrum is still intact. This noise only is applied to modes $>$ 10 $\mu$m so as to not introduce new information into the image. Applying this method to the \textit{DRACO} velocity output results in velocity spectra and 2-D images that can be quantitatively and qualitatively compared to the experimental data respectively.

The 1-D velocity perturbations for one-beam simulations are presented in Fig. 5(a). This velocity is extrapolated to 2-D and placed on the same color scale as Fig. 2(a). It should be noted that since the resolution of the simulations (10 $\mu$m) is less than that of the diagnostic (2 $\mu$m), modes smaller than 10 $\mu$m do not exist in the extrapolated simulation data, while they do in the experimental data. This means that the simulations will never be able to capture the finer structure seen in the experimental data. When comparing the amplitude of the velocity modulations between \textit{DRACO} and the experiment for modes between 10 and 100 $\mu$m in size, it is evident that current simulations dramatically underestimate the level of imprint for these modes. Fig. 5(e) compares the velocity spectrum between the experiment and \textit{DRACO}; \textit{DRACO} underestimates the amplitude of the velocity spectrum by a factor of 3. One-beam \textit{DRACO} simulations return a velocity $\sigma_\mathrm{rms}$ of 385 m/s, while the two-beam \textit{DRACO} case returns 330 m/s, both significantly below the experimental results of 1030 and 550 m/s, respectively. 

One factor that could account for \textit{DRACO} underestimating imprint is that it lacks the multiphoton ionization (MPI) process for the initial formation of plasma \cite{Bourgeade}. MPI of a material occurs with the simultaneous absorption of several photons with energy less than the material's ionization threshold ~\cite{Mainfray,Voronov}. Without the inclusion of MPI the pressure exerted by a beam profile can typically be described by $\delta P/P \ \propto \ \delta n_e/n_e \ \propto \ \delta I/I$ for a fully ionized plasma, where $P$ is the ablation pressure imposed by the laser intensity $I$. At the onset of laser irradiation, the polystyrene is dielectric with no conduction band (free) electrons. MPI is expected to generate free electrons followed by avalanche ionization that eventually drives the plasma-generation process. For such a case, assuming that electron density $n_e$ has not yet saturated, the pressure should scale with laser intensity as
\begin{equation}
\frac{\delta P}{P} \propto \frac{\delta n_e[I^\alpha]}{n_e[I^\alpha]} \sim \alpha \frac{\delta I}{I}.
\end{equation}
Here $\alpha$ is the number of photons needed to pump the valence electrons to the conduction band ~\cite{Mezel,Temnov}. For polystyrene, $\alpha = 2$ is a reasonable estimate. To test this, \textit{DRACO} simulations were repeated with the laser modulation amplitude increased by a factor of 2 to account for MPI. This resulted in a large increase in the amplitude of the velocity perturbations as shown in Fig. 5(b) and (d). The velocity spectrum [Fig. 5(e)] is in much better agreement with the experimental data. The simulated velocity front now has a $\sigma_\mathrm{rms} \approx 1100$ m/s, which is much more comparable with the experiment's 1030 m/s than the 330 m/s simulated using the DRACO approximation without MPI.

When comparing the velocity spectra it appears that the simulated spectrum is much more modulated than the experimental one. This can be attributed primarily to the azimuthal averaging of the velocity spectrum in the experiment across two dimensions, while the spectrum from \textit{DRACO} contains data across only one dimension. The azimuthal averaging process tends to significantly smooth the spectrum in the experimental case. One can see in Fig. 2(c) that in the single beam case there is large uncertainty in the velocity spectrum for larger wavelength modes. This is due to sampling: a 1-D lineout taken of the experimental data containing a large feature translates it to a significantly higher amplitude of a large wavelength mode. When 1-D lineouts are constructed and averaged across an entire image this mode spike becomes less significant and a smoother curve is generated. The simulations only produce a single lineout of data and so the smoothing benefits of the averaging process are lost.
\\
\section{Conclusions}

In summary, 2-D VISAR images of shock velocity perturbations produced by laser intensity modulation have provided the first direct measurement of laser imprint. Experiments showed that using two and five overlapping beams reduced imprint-induced shock velocity modulations across all spatial modes when compared to the single beam case. However, these initial experiments revealed a deficiency in current laser-imprint simulations, which underestimated induced velocity perturbations by a factor of three. By accounting for microphysics processes during the initial plasma formation, such as multiphoton ionization, a more realistic model of laser imprint for laser direct drive inertial confinement fusion implosions has been developed.


\begin{acknowledgments}
This material is based upon work supported by the Department of Energy National Nuclear Security Administration under Award Number DE-NA0003856, the University of Rochester, and the New York State Energy Research and Development Authority. This work was performed under the auspices of the U.S. Department of Energy by Lawrence Livermore National Laboratory under Contract DE-AC52-07NA27344.

This report was prepared as an account of work sponsored by an agency of the U.S. Government. Neither the U.S. Government nor any agency thereof, nor any of their employees, makes any warranty, express or implied, or assumes any legal liability or responsibility for the accuracy, completeness, or usefulness of any information, apparatus, product, or process disclosed, or represents that its use would not infringe privately owned rights. Reference herein to any specific commercial product, process, or service by trade name, trademark, manufacturer, or otherwise does not necessarily constitute or imply its endorsement, recommendation, or favoring by the U.S. Government or any agency thereof. The views and opinions of authors expressed herein do not necessarily state or reflect those of the U.S. Government or any agency thereof.	
\end{acknowledgments}


\begin{thebibliography}{9}\label{sec:TeXbooks}%
	
\bibitem{Craxton}
R. S. Craxton, K. S. Anderson, T. R. Boehly, V. N. Goncharov, D. R. Harding, J. P. Knauer, R. L. McCrory, P. W. McKenty, D. D. Meyerhofer, J. F. Myatt \textit{et al.}, Phys. Plasmas \textbf{22}, 110501 (2015).
%

\bibitem{Bodner}
S. E. Bodner, Phys. Rev. Lett. \textbf{33}, 761 (1974).
%

\bibitem{Bourgeade}
A. Bourgeade and G. Duchateau, Phys. Rev. E \textbf{85}, 056403 (2012).
%
	
\bibitem{Remington}
B. A. Remington, S. V. Weber, M. M. Marinak, S. W. Haan, J. D. Kilkenny, R. Wallace, and G. Dimonte, Phys. Rev. Lett. \textbf{73}, 545 (1994).
%
	
\bibitem{Glendinning}
S. G. Glendinning, S. N. Dixit, B. A. Hammel, D. H. Kalantar, M. H. Key, J. D. Kilkenny, J. P. Knauer, D. M. Pennington, B. A. Remington, R. J. Wallace \textit{et al.}, Phys. Rev. Lett. \textbf{78}, 3318 (1997).
%
	
\bibitem{Azechi}
H. Azechi, M. Nakai, K. Shigemori, N. Miyanaga, H. Shiraga, H. Nishimura, M. Honda, R. Ishizaki, J. G. Wouchuk, H. Takabe \textit{et al.}, Phys. Plasmas \textbf{4}, 4079 (1997).
%
	
\bibitem{Azechi2}
H. Azechi, T. Sakaiya, S. Fujioka, Y. Tamari, K. Otani, K. Shigemori, M. Nakai, H. Shiraga, N. Miyanaga, and K. Mima, Phys. Rev. Lett. \textbf{98}, 045002 (2007).
%
	
\bibitem{Cherfils}
C. Cherfils, S. G. Glendinning, D. Galmiche, B. A. Remington, A. L. Richard, S. Haan, R. Wallace, N. Dague, and D. H. Kalantar, Phys. Rev. Lett. \textbf{83}, 5507 (1999).
%
	
\bibitem{Smalyuk}
V. A. Smalyuk, T. R. Boehly, D. K. Bradley, V. N. Goncharov, J. A. Delettrez, J. P. Knauer, D. D. Meyerhofer, D. Oron, and D. Shvarts, Phys. Rev. Lett. \textbf{81}, 5342 (1998).
%
	
\bibitem{Smalyuk2}
V. A. Smalyuk, O. Sadot, J. A. Delettrez, D. D. Meyerhofer, S. P. Regan, and T. C. Sangster, Phys. Rev. Lett. \textbf{95}, 215001 (2005).
%

\bibitem{Hu2}
S. X. Hu, D. T. Michel, A. K. Davis, R. Betti, P. B. Radha, E. M. Campbell, D. H. Froula, and C. Stoeckl, Phys. Plasmas \textbf{23}, 102701 (2016).
%

\bibitem{Michel}
D. T. Michel, S. X. Hu, A. K. Davis, V. Yu. Glebov, V. N. Goncharov, I. V. Igumenshchev, P. B. Radha, C. Stoeckl, and D. H. Froula, Phys. Rev. E \textbf{95}, 051202(R) (2017).
%
\bibitem{Hu4}
S. X. Hu, V. N. Goncharov, P. B. Radha, J. A. Marozas, S. Skupsky, T. R. Boehly, T. C. Sangster, D. D. Meyerhofer, and R. L. McCrory, Phys. Plasmas \textbf{17}, 102706 (2010).
%

\bibitem{Carr}
C. W. Carr, H. B. Radousky, A. M. Rubenchik, M. D. Feit, and S. G. Demos, Phys. Rev. Lett. \textbf{92}, 087401 (2004).
%

\bibitem{Kato}
Y. Kato, K. Mima, N. Miyanaga, S. Arinaga, Y. Kitagawa, M. Nakatsuka, and C. Yamanaka, Phys. Rev. Lett. \textbf{53}, 1057 (1984).
%
	
\bibitem{Lin}
Y. Lin, T. J. Kessler, and G. N. Lawrence, Opt. Lett. \textbf{20}, 764 (1995).
%
	
\bibitem{Skupsky}
S. Skupsky, R. W. Short, T. Kessler, R. S. Craxton, S. Letzring, and J. W. Soures, J. Appl. Phys. \textbf{66}, 3456 (1989).
%
	
\bibitem{Regan}
S. P. Regan, J. A. Marozas, J. H. Kelly, T. R. Boehly, W. R. Donaldson, P. A. Jaanimagi, R. L. Keck, T. J. Kessler, D. D. Meyerhofer, W. Seka \textit{et al.}, J. Opt. Soc. Am. B \textbf{17}, 1483 (2000).
%
	
\bibitem{Regan2}
S. P. Regan, J. A. Marozas, R. S. Craxton, J. H. Kelly, W. R. Donaldson, P. A. Jaanimagi, D. Jacobs-Perkins, R. L. Keck, T. J. Kessler, D. D. Meyerhofer \textit{et al.}, J. Opt. Soc. Am. B \textbf{22}, 998 (2005).
%

\bibitem{Lehmberg}
R. H. Lehmberg, A. J. Schmitt, and S. E. Bodner, J. Appl. Phys. \textbf{62}, 2680 (1987).
%

\bibitem{Boehly2}
T. R. Boehly, V. A. Smalyuk, D. D. Meyerhofer, J. P. Knauer, D. K. Bradley, R. S. Craxton, M. J. Guardalben, S. Skupsky, and T. J. Kessler, J. Appl. Phys. \textbf{85}, 3444 (1999).
%

\bibitem{Lehmberg2}
R. H. Lehmberg and S. P. Obenschain, Opt. Commun. \textbf{46}, 27–31 (1983);
%

\bibitem{Deniz}
A. V. Deniz, T. Lehecka, R. H. Lehmberg, and S. P. Obenschain, Opt. Commun. \textbf{147}, 402–410 (1998).
%

\bibitem{Kalantar}
D. H. Kalantar, M. H. Key, L. B. Da Silva, S. G. Glendinning, B. A. Remington, J. E. Rothenberg, F. Weber, S. V. Weber, E. Wolfrum, N. S. Kim \textit{et al.}, Phys. Plasmas \textbf{4}, 1985 (1997).
%
	
\bibitem{Kalantar2}
D. H. Kalantar, T. W. Barbee, Jr., L. B. Da Silva, S. G. Glendinning, F. Weber, M. H. Key, and J. P. Knauer, Rev. Sci. Instrum. \textbf{67}, 781 (1996).
%
	
\bibitem{Kalantar3}
D. H. Kalantar, L. B. Da Silva, S. G. Glendinning, B. A. Remington, F. Weber, S. V. Weber, A. Demir, M. H. Key, N. S. Kim, J. P. Knauer \textit{et al.}, Rev. Sci. Instrum. \textbf{68}, 802 (1997).
%
	
\bibitem{Depierreux}
S. Depierreux, C. Labaune, D. T. Michel, C. Stenz, P. Nicolai, M. Grech, G. Riazuelo, S. Weber, C. Riconda, V. T. Tikhonchuk \textit{et al.}, Phys. Rev. Lett. \textbf{102}, 195005 (2009).
%
	
\bibitem{Obenschain}
S. P. Obenschain, D. G. Colombant, M. Karasik, C. J. Pawley, V. Serlin, A. J. Schmitt, J. L. Weaver, J. H. Gardner, L. Phillips, Y. Aglitskiy \textit{et al.}, Phys. Plasmas \textbf{9}, 2234 (2002).
%
	
\bibitem{Karasik}
M. Karasik, J. L. Weaver, Y. Aglitskiy, J. Oh, and S. P. Obenschain, Phys. Rev. Lett. \textbf{114}, 085001 (2015).
%
	
\bibitem{Mostovych}
A. N. Mostovych, D. G. Colombant, M. Karasik, J. P. Knauer, A. J. Schmitt, and J. L. Weaver, Phys. Rev. Lett. \textbf{100}, 075002 (2008).
%
	
\bibitem{Fujioka}
S. Fujioka, A. Sunahara, K. Nishihara, N. Ohnishi, T. Johzaki, H. Shiraga, K. Shigemori, M. Nakai, T. Ikegawa, M. Murakami \textit{et al.}, Phys. Rev. Lett. \textbf{92}, 195001 (2004).
%
	
\bibitem{Hu}
S. X. Hu, G. Fiksel, V. N. Goncharov, S. Skupsky, D. D. Meyerhofer, and V. A. Smalyuk, Phys. Rev. Lett. \textbf{108}, 195003 (2012).
%
	
\bibitem{Hu3}
S. X. Hu, W. Theobald, P. B. Radha, J. L. Peebles, S. P. Regan, A. Nikroo, M. J. Bonino, D. R. Harding, V. N. Goncharov, N. Petta \textit{et al.}, Phys. Plasma \textbf{25}, 082710 (2018).
%
	
\bibitem{Olazabal-Loumé}
M. Olazabal-Loumé, Ph Nicolaï, G. Riazuelo, M. M. Grech, J. Breil, S. Fujioka, A. Sunahara, A. Borisenko, and V. T. Tikhonchuk, New J. Phys. \textbf{15}, 085033 (2013).
%
	
\bibitem{Masse}
L. Masse, Phys. Rev. Lett. \textbf{98}, 245001 (2007).
%

	
\bibitem{Boehly}
T. R. Boehly, D. L. Brown, R. S. Craxton, R. L. Keck, J. P. Knauer, J. H. Kelly, T. J. Kessler, S. A. Kumpan, S. J. Loucks, S. A. Letzring \textit{et al.}, Opt. Commun. \textbf{133}, 495 (1997).
%
	
\bibitem{Celliers}
P. M. Celliers, D. J. Erskine, C. M. Sorce, D. G. Braun, O. L. Landen, and G. W. Collins, Rev. Sci. Instrum. \textbf{81}, 035101 (2010).
%
	
\bibitem{Hemsing}
W. F. Hemsing, Rev. Sci. Instrum. \textbf{50}, 73 (1979).
%
	
\bibitem{Erskine}
D. J. Erskine, R. F. Smith, C. A. Bolme, P. M. Celliers, and G. W. Collins, Rev. Sci. Instrum. \textbf{83}, 043116 (2012).
%
	
\bibitem{Erskine2}
D. J. Erskine, R. F. Smith, C. Bolme, S. Ali, P. M. Celliers, and G. W. Collins, J. Phys.: Conf. Ser. \textbf{500}, 142013 (2014).
%
	
\bibitem{Ghiglia}
D. C. Ghiglia and L. A. Romero, J. Opt. Soc. Am. A \textbf{11}, 107 (1994).
%
	
\bibitem{Ghiglia2}
D. C. Ghiglia and L. A. Romero, J. Opt. Soc. Am. A \textbf{13}, 1999 (1996).
%
	
\bibitem{Theobald}
W. Theobald, J. E. Miller, T. R. Boehly, E. Vianello, D. D. Meyerhofer, T. C. Sangster, J. Eggert, and P. M. Celliers, Phys. Plasmas \textbf{13}, 122702 (2006).
%
	
\bibitem{Smalyuk3}
V. A. Smalyuk, V. N. Goncharov, T. R. Boehly, J. A. Delettrez, D. Y. Li, J. A. Marozas, A. V. Maximov, D. D. Meyerhofer, S. P. Regan, and T. C. Sangster, Phys. Plasmas \textbf{12}, 072703 (2005).
%
	
\bibitem{Radha}
P. B. Radha, T. J. B. Collins, J. A. Delettrez, Y. Elbaz, R. Epstein, V. Yu. Glebov, V. N. Goncharov, R. L. Keck, J. P. Knauer, J. A. Marozas \textit{et al.}, Phys. Plasmas \textbf{12}, 056307 (2005).
%

\bibitem{Hu5}
S. X. Hu, V. A. Smalyuk, V. N. Goncharov, J. P. Knauer, P. B. Radha, I. V. Igumenshchev, J. A. Marozas, C. Stoeckl, B. Yaakobi, D. Shvarts, T. C. Sangster, P. W. McKenty, D. D. Meyerhofer, S. Skupsky, and R. L. McCrory, Phys. Rev. Lett. \textbf{100}, 185003 (2008)
%
\bibitem{Epstein}
R. Epstein, J. Appl. Phys. \textbf{82}, 2123 (1997).
%

\bibitem{Mainfray}
G. Mainfray and C. Manus, Rep. Prog. Phys. \textbf{54}, 1333 (1991).
%
	
\bibitem{Voronov}
G. S. Voronov and N. B. Delone, Sov. Phys. JETP \textbf{23}, 54 (1966).
%

\bibitem{Mezel}
C. M\'ezel, G. Duchateau, G. Geneste, and B. Siberchicot, J. Phys.: Condens. Matter \textbf{25}, 235501 (2013)
%

\bibitem{Temnov}
V. V. Temnov, K. Sokolowski-Tinten, P. Zhou, A. El-Khamhawy, and D. von der Linde, Phys. Rev. Lett. \textbf{97}, 237403 (2006).
	
\end{thebibliography}
\end{document}